# A General Optimal Control Model of Human Movement Patterns II: Rapid, Targeted Hand Movements (Fitts' Law)

**Stuart Hagler**

Oregon Health & Science University

Portland, OR, USA

haglers@ohsu.edu

**Abstract:** Rapid, targeted hand movements exhibit a regular movement pattern described by Fitts' law. We develop a model of these movements in which this movement pattern results from an optimal control model describing rapid hand movements and a utility model describing the speed/accuracy trade-off between moving the hand rapidly to the target and hitting the target accurately. The optimal control model is constructed using principled approach in which we forbid the muscle forces to exhibit any discontinuities and require the cost to be expressed in terms of a psychophysical representation of the movement. This yields a yank-control or jerk-control model of the movement which exhibits two constants of the motion that are closely related to the energy and momentum in classical mechanics. We force the optimal control model to obey Fitts' law by requiring a particular relationship hold between the constants of the motion and the size of the target and show that the resulting model compares well to a standard expression of Fitts' law obtained empirically using observations of computer mouse movements. We then proceed to further show how this relationship may be obtained as the result of a simple models of the movement accuracy and the speed/accuracy trade-off. We use the movement accuracy model to analyze observed differences in computer mouse movement patterns between older adults with mild cognitive impairment and intact older adults. We conclude by looking at how a subject might carry out in practice the optimization implicit in resolving the speed/accuracy trade-off in our model.

## 1 Introduction

Rapid, targeted hand movements, that is, rapid movements of the hand into a specified target region, exhibit a regular movement pattern described by Fitts' law. [1-4] This law provides an implicit account of the speed/accuracy trade-off in a rapid, targeted hand movement. That is, it captures a subject's resolution to the problem of whether to move slowly to a target but hit it with very high accuracy, or to move quickly to a target but with much lower accuracy, or to choose some intermediate combination of speed and accuracy. We have argued [5] that regular movement patterns arise from regularities in how subjects selects a particular movement from the range of possible movements. In [6], we constructed a general optimal control model to account for regularities in the human movement trajectories, and applied this model to the description of walking gait. In the present paper, we apply this model to the description of rapid, targeted hand movements. When we described walking gait in [6], we did not consider movement accuracy. We can argue that this was a reasonable assumption in the case of walking gait as the accuracy there can safely be ignored because even with low movement accuracies a subject can still execute a walking gait. In the case of rapid, targeted hand movements, the targets are sufficiently small that the movement accuracy must be much higher and must be included in the model. We therefore find that, for the present problem, we must extend the general optimal control in [6] to include an account of movement accuracy and the speed/accuracy trade-off.

A number of models for Fitts' law have been proposed including control theoretic models (e.g. [7-13]), as well as an earlier version of the present treatment [14]). The approach we adopt is to begin with the general optimal control model proposed in [6] and used there to describe walking gait and to extend it to include a models of accuracy and of the speed/accuracy trade-off. Although implicit in [6], we make it more explicit here that the optimal control model is constructed by requiring it satisfy specific physical principles. The first principle requires that the forces generated by the muscles be continuous throughout time having no discontinuous jumps in the applied muscle forces. In particular, we have in mind that no discontinuities occur at the instants in time at which the movement begins or ends (we have discussed this in detail in [6, 14]). The adoption of this principle results in a model in which a movement cost related to the yank (i.e. the time-derivative of the force) or the jerk (i.e. the time-derivative of the acceleration) is minimized. Jerk minimization as a general principle for the description of human movements has been proposed by several researchers. [15-20] The second principle requires that the movement cost be expressed as a psychophysical model in which any relevant mechanical quantities appear in terms of psychophysical functions which describe how the mechanical quantities are perceived by the subject. The movement cost thus becomes a description of how the subject perceives the movement and weighs the various perceptions of the values of mechanical quantities.

While rapid, targeted hand movements are of basic scientific interest for understanding human motor control, they have also been studied for practical applications, particularly movements of a computer mouse during routine computer usage. Computer mouse movements have been shown to obey Fitts' law. [21, 22] Indeed, one of the early applications of Fitts' law was to the design of graphical user interfaces to maximize ease-of-use, [21] and computer mouse movements have been used in techniques for verifying the identity of a computer user. [23-27] It has also been used to capture average movement times of the computer mouse to better understand the timing of other cognitive activities during the play of simple mouse-controlled computer games. [28, 29] More generally, mouse movement patterns have been shown to be able to distinguish subjects with mild cognitive impairment (MCI) from intact subjects, [30] and alone [31] and when combined with keystrokes have formed the basis of proposed techniques for detecting of MCI using interactions with a computer. [32, 33] The measurement of emotionalist states of computer users so that an intelligent computer system may better meet a user's needs. [34]

This paper is structured as follows. We first provide a brief treatment of Fitts' law (Sec. 2). We next provide some definitions and summarize the optimal control approach to modeling movements that we developed in [6] (Sec. 3). We then construct the optimal control model for rapid hand movements (Sec. 4). We finally extend the optimal control model for rapid hand movements to the description of rapid, targeted hand movements by showing what relationships the model must satisfy for it to be forced to obey Fitts' law and how these relationships arise from simple models of the movement accuracy and the speed/accuracy trade-off. We compare the resulting model of Fitts' law to empirical data for computer mouse movements reported by MacKenzie, [22] and we use the resulting model of movement accuracy to analyze the differences in computer mouse movement patterns for older adults with MCI and intact older adults reported by Seelye et al. [30] We consider how a subject may perform in practice the optimization implicit in the model of the speed/accuracy trade-off. (Sec. 5).

**2 Fitts' Law**

Fitts' law describes rapid, targeted hand movements by relating the total movement time $T$ required to complete the movement to the distance $D$ moved from the initial position to the center of the target region and the width $W$ of the target region along the direction of motion of the hand. In addition to Fitts' original formulation, [1, 2] a number of alternative mathematical forms for Fitts' law have been proposed (see e.g., [4, 35, 36]). The form of Fitts' law that we adopt is taken from [4, 21, 22]; this form of Fitts' law gives the relationship:

$$T = a + b \log_2 \bigl( D / W + 1 \bigr). \tag{1}$$

The *index of difficulty* $log_2(D/W + 1)$ provides a measurement of the amount of information that must be processed in the parallel cognitive process; as the amount of information to be processed increases, the movement slows (i.e. the movement time increases) to accommodate the extra time needed to process the additional information; it is measured in bits. The constant $b$ gives the time the cognitive process takes to process each bit of information.

In an experiment designed to measure the movement time $T$ for a rapid, targeted hand movement, we start the clock when the subject is presented with a target. The subject then performs the necessary planning and executes the movement. Following [28] and [37, 38], we analyze the movement time $T$ using an additive model consisting of two stages: (i) a planning stage requiring a time $T_P(D/W)$, and (ii) a motor stage requiring a time $T_M(D/W)$ to complete. The total time $T$ to move to the target is then:

$$T = T_P\bigl(D / W\bigr) + T_M\bigl(D / W\bigr). \tag{2}$$

We have written $T_P(D/W)$ and $T_M(D/W)$ as functions of $D/W$ since we interpret (1) as saying that neither the movement distance $D$ moved nor the target width $W$ occur alone in the model but only in the ratio $D/W$. For simplicity, we assume that the dependence of the planning stage on $D/W$ is much smaller than that of the motor stage; thus:

$$T \approx \tau_P + T_M\bigl(D / W\bigr). \tag{3}$$

**3 Optimal Control Model**

We now briefly develop terminology and notation as well as provide an outline of the mathematical apparatus that we need to develop the optimal control model of rapid hand movements in Sec. 4. The mathematical apparatus has been developed in more detail in [6, 14]. We find that it is mathematically convenient in the subsequent analysis to take movements taking a time $T$ to begin at time $-T/2$ and end at time $T/2$.

*3.1 Terminology & Notation*

For a position variable $x(t)$, the first- and second-order time-derivatives are the velocity $v(t) = \dot{x}(t)$ and the acceleration $a(t) = \ddot{x}(t)$. We follow [14] and denote the higher order time-derivatives as the *jerk* $\dddot{x}(t)$, the *snap* $\ddddot{x}(t)$, the *crackle* $\dddddot{x}(t)$, and the *pop* $\ddddddot{x}(t)$. For a force variable $F(t)$ (e.g. example representing the net force of the muscles acting on a point on the body as a function of time), we denote the first-order time-derivative as the *yank* $\dot{F}(t)$. The *impulse* is the time integral of the force from the beginning of the movement is $I(t) = \int_{-T/2}^{t} F(t') \, dt'$. For movements beginning at rest, the impulse is simply the momentum at time $t$. We denote an ordinary function as $\xi(\cdot)$ using parentheses, and a functional (a function of a function) as $\xi[\cdot]$ using brackets. We denote the set of all values $\xi_n$ as $\{\xi_n\}$. A function $\xi(t)$ is *odd* if $\xi(t)$ $= -\xi(-t)$, and it is *even* if $\xi(t) = \xi(-t)$.

*3.2 Segment, Metabolic Energy, & Perceived Muscle Force Models*

We model the skeleton of the human body as in [5, 6] using a segment model consisting of a system of N ("nu") segments attached at $N$ joints. A movement is described by a set of trajectories $x(t)$ of points on the body and the rest of the body is constrained to move sensibly given the motion of this specified set of points. Associated with each controlled point on the body is an effective mass $m$, a trajectory $x(t)$, and a net muscle force $F(t)$ that determines the acceleration of that point according to Newton's second law using the effective mass. In the model that we use to describe rapid hand movements, we only allow for a single control point in the hand with the rest of the body held still. We develop the remainder of the model with a single control point in mind.

*3.3 Cost Functional*

The movement of the controlled point on the body is governed by a cost functional $J[F,\acute{F}]$ of the form:

$$J[F,\dot{F}] = \int_{-T/2}^{T/2} \left( \alpha \dot{F}^2 + \varepsilon F^2 \right) dt. \tag{4}$$

As we will only be interested in rapid hand movements in one dimension, we have written (4) in a one-dimensional form. The yank term in $\acute{F}^2$ causes the model to obey the principle that there should be no discontinuities in the force (as we have shown in [6, 14]), while the force term in $F^2$ indicates that we are including the muscle force in a form that satisfies the principle that the model should include the perceived force of the muscle, in this case in a form consistent with Stevens' power law [39, 40] (as we have shown in [5, 6]). This model does not fully accord with the principles that we gave in Sec. 1 as we have failed provide a psychophysical account of the yank term and it is unclear to what extent a subject perceives the yank of a movement. However, we show in Sec. 4.4.2 that the yank for the optimal movement which satisfies (4) may be expressed in terms of the velocity. This presents the possibility for giving a psychophysical account of the yank term in terms of the velocity assuming a psychophysical account of the perception of the velocity of a movement can be made (as seems likely). We have argued in [6] that we can rewrite the yank-control cost functional $J[F,\acute{F}]$ in terms of the set of functions $x$, $\dot{x}$, $\ddot{x}$, and $\dddot{x}$ as the jerk-control cost functional:

$$J[x,\dot{x},\ddot{x},\dddot{x}] = \int_{-T/2}^{T/2} L\left(x,\dot{x},\ddot{x},\dddot{x}\right) dt. \tag{5}$$

We note that the highest-order derivative of the trajectory is the jerk $\dddot{x}$, thus we are asserting in (5) that the yank-control model in (4) has become a jerk-control model.

*3.4 Lagrangian Method of Solving the Optimal Control Problem*

In the Lagrangian method of solving the optimal control problem of finding the trajectory $x(t)$ that minimizes the value of the cost functional in (5), the optimal trajectory satisfies the differential equation:

$$\frac{\partial L}{\partial x} - \frac{d}{dt}\frac{\partial L}{\partial \dot{x}} + \frac{d^2}{dt^2}\frac{\partial L}{\partial \ddot{x}} - \frac{d^3}{dt^3}\frac{\partial L}{\partial \dddot{x}} = 0. \tag{6}$$

Here $L$ is the integrand $L(x,\dot{x},\ddot{x},\dddot{x})$ in (5) and is known as the *Lagrangian* of the system.

*3.5 Hamiltonian Methods of Solving the Optimal Control Problem*

The Hamiltonian method of solving the optimal control problem begins by defining the *generalized coordinate vector $Q$*, *control $u$*, and *generalized momentum vector $P$* as:

$$\begin{aligned} Q^{\mathrm{T}} &= \begin{bmatrix} x, & \dot{x}, & \ddot{x} \end{bmatrix}, \\ u &= \dddot{x}, \\ P^{\mathrm{T}} &= \begin{bmatrix} p_1, & p_2, & p_3 \end{bmatrix}. \end{aligned} \qquad (7)$$

The *Hamiltonian $H$* is obtained by taking the Legendre transform of the Lagrangian:

$$H = P^{\mathrm{T}} \dot{Q} - L. \qquad (8)$$

The trajectory $x(t)$ that minimizes the value of the cost functional in (5) satisfies the system of differential equations given by:

$$\begin{aligned} \dot{Q} &= \partial H / \partial P, & (A) \\ \dot{P} &= -\partial H / \partial Q, & (B) \\ \partial H / \partial u &= 0. & (C) \end{aligned} \qquad (9)$$

The Hamiltonian takes a constant value when calculated along the optimal trajectory (i.e. the trajectory that solves the systems in (6) and (9)); we call this constant value the *generalized energy $\Psi$*. Along the optimal trajectory we find:

$$H(t) = \Psi. \qquad (10)$$

**4 Rapid Hand Movements**

Rapid hand movements are movements where the hand rapidly traverses a specified movement distance in a specified direction with low accuracy. In this case, the accuracy is assumed to be sufficiently low that we can neglect how it affects how the movement is executed. We now construct an optimal control model of rapid hand movements using the approach outlined in Sec. 3. We proceed by first providing a useful anthropometric value. We then model hand movements using a segment model consisting of a single segment with the body at one end and with a "hand" at the other end containing all the mass of the hand and arm. We construct a yank-control cost functional in terms of the muscle force trajectories, which we rewrite as a jerk-control cost functional in terms of the body segment trajectories. We then find the optimal trajectories of the body using the Lagrangian method of solving the optimal control problem. We finally calculate the generalized momentum and generalized energy for rapid hand movements using the Hamiltonian method of solving the optimal control problem, and identify two constants of the motion closely related to the energy and momentum in classical mechanics.

*4.1 An Anthropometric Value*

A subject with mass $M$ has a mass in the hand and arm of about $m \approx 0.05 M$. [41]

*4.2 Rapid Hand Movement Model*

We model the body using a one-segment model with one segment for the arm being moved — the arm is straight, does not bend at the elbow, but can change length. The mass of the body is placed in the *body* which is the point at one end of the segment that we assume never moves; the mass of the hand and arm is placed in the *hand* at the other end of the segment. We assume the movement is such that the hand only moves within the comfortable range of motion of the arm. The hand begins the movement at rest in some position and ends the movement at rest in another. We can reasonably expect that the hand moves along a straight line between the beginning and ending positions so that we need only solve for the trajectory along that straight line. For movements in which the hand is held up throughout the movement, while gravity affects the metabolic energy expended during the movement it does not affect the optimal trajectory, thus we may neglect gravity. Alternatively, for movements in which the hand rests on a table throughout the movement, we may neglect gravity due to the presence of the table. For convenience, and since we will be examining mouse movements in the following section, we examine the movement of the hand along the surface of a frictionless table. This situation is illustrated in Fig. 1.

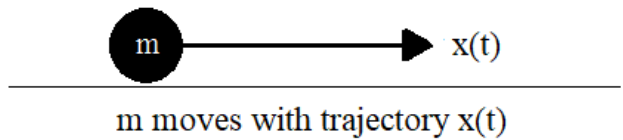

**Figure 1.** The motion of the hand along the table during a rapid movement. The hand moves with trajectory x(t) in a straight line along the surface of the table.

The hand moves according to the net force $F(t)$ acting on it due to the action of muscles. The trajectory $x(t)$ of the hand then trivially satisfies Newton's second law:

$$F = m\ddot{x}. \tag{11}$$

*4.3 Cost Functional*

The cost functional for rapid hand movements is just that given in (4); it is:

$$J = \int_{-T/2}^{T/2} \left( \alpha_h \dot{F}^2 + \varepsilon_h F^2 \right) dt. \tag{12}$$

For convenience, we define the frequency parameter value $\omega_h$ given by:

$$\omega_h^2 = \varepsilon_h / \alpha_h. \tag{13}$$

Using this frequency parameter, the cost functional $J$ may be rewritten in terms of the trajectory $x(t)$ of the hand as:

$$J = \alpha_h m^2 \int_{-T/2}^{T/2} \left( \dddot{x}^2 + \omega_h^2 \ddot{x}^2 \right) dt. \tag{14}$$

*4.4 Optimal Trajectory*

The optimal trajectory minimizes the cost functional $J$ in (14). We first calculate the optimal trajectory using the Lagrangian method of solving the optimal control problem. This gives a differential equation the solution of which is the optimal trajectory of the hand. We then calculate the generalized momentum vector and generalized energy using the Hamiltonian method of solving the optimal control problem.

*4.4.1 Lagrangian Method*

The Lagrangian $L$ governing the motion of the hand is the integrand of the cost functional $J$ in (14); that is:

$$L = \alpha_h m^2 \left( \dddot{x}^2 + \omega_h^2 \ddot{x}^2 \right). \tag{15}$$

We obtain the equation of motion from the Lagrangian by taking the partial derivatives of $L$ using the various time-derivatives of $x(t)$ according to (6); it is:

$$\ddddot{\ddot{x}} - \omega_h^2 \ddddot{x} = 0. \tag{16}$$

The trajectory $x(t)$ that solves (16) is:

$$\begin{aligned} x(t) &= c_1 + c_2 t + c_3 t^2 + c_4 t^3 \\ &\quad + c_5 \sinh(\omega_h t) + c_6 \cosh(\omega_h t). \end{aligned} \tag{17}$$

The trajectory in (17) is parameterized by six parameters – $c_1, \ldots, c_6$. We observe that the trajectory $x(t)$ consists of four terms in $t$ related to the motion of the hand as a point particle on the table and two terms in $\omega_h t$ related to the force of muscle acting on the hand. We may understand the two terms in $\omega_h t$ as providing a description of how the acceleration $\ddot{x}(t)$ is controlled by the muscles.

The hand must traverse the distance $D$ separating the initial and final positions in the movement time $T$. It must begin and end each movement resting on the table with no velocity or acceleration. The initial and final conditions of the motion are:

$$\begin{aligned} x(-T/2) &= -D/2, & x(T/2) &= D/2, \\ \dot{x}(-T/2) &= 0, & \dot{x}(T/2) &= 0, \\ \ddot{x}(-T/2) &= 0, & \ddot{x}(T/2) &= 0. \end{aligned} \tag{18}$$

In this case, the movement of the hand is symmetric and only three of the six terms in the optimal trajectory $x(t)$ of the hand in (17) remain (have non-zero parameter values), namely those that are odd in time on the interval $-T/2 \leq t \leq T/2$. This gives an optimal trajectory $x(t)$ of the hand of the form:

$$x(t) = C_1 t + C_2 t^3 + C_3 \sinh(\omega_h t). \tag{19}$$

It is convenient at this point to introduce a constant parameter $p_1$ defined to satisfy:

$$C_2 = -p_1 / 12\varepsilon_h m^2. \tag{20}$$

The constant parameter $p_1$ is a term in the generalized momentum and will be defined in Sec. 4.4.2, but at this point is a convenience to simplify our expressions. Combining (19) and (20), we find:

$$x(t) = C_1 t - \left(p_1 / 12\varepsilon_h m^2\right) t^3 + C_3 \sinh(\omega_h t). \tag{21}$$

We may now solve the optimal trajectory in (21) given the initial and final conditions in (18). We calculate the values for $C_1$ and $C_3$ in (21) in Appendix 1; they are:

$$\begin{aligned}
C_1 &= \left(1 / 4\alpha_h \omega_h^4 m^2\right)\left(\left(\omega_h T / 2\right)^2 - 2\left(\omega_h T / 2\right)\coth\left(\omega_h T / 2\right)\right) p_1, \\
C_3 &= \left(1 / 2\alpha_h \omega_h^5 m^2\right)\left(\left(\omega_h T / 2\right) / \sinh\left(\omega_h T / 2\right)\right) p_1.
\end{aligned} \tag{22}$$

We also calculate the value for $p_1$ in (21) in Appendix 1; it is:

$$\begin{aligned}
p_1 &= \left(\alpha_h \omega_h^5 m^2\right) D / \pi\left(\omega_h T / 2\right), \\
\pi\left(\omega_h T / 2\right) &= -\left(\omega_h T / 2\right)^2 \left(\coth\left(\omega_h T / 2\right) - \left(\omega_h T / 2\right)^{-1} - \left(\omega_h T / 2\right) / 3\right).
\end{aligned} \tag{23}$$

Using (11), we find that the force $F$ generated on the hand by the arm is:

$$F = -\left(p_1 / 2\varepsilon_h m\right) t + C_3 \omega_h^2 m \sinh(\omega_h t). \tag{24}$$

*4.4.2 Hamiltonian Method*

The Hamiltonian $H$ governing the motion of the hand is given by:

$$H = P^{\mathrm{T}} \dot{Q} - L. \tag{25}$$

The generalized coordinates vector $Q$ and the control $u$ are:

$$\begin{aligned}
Q^{\mathrm{T}} &= \begin{bmatrix} x, & \dot{x}, & \ddot{x} \end{bmatrix}, \\
u &= \dddot{x}.
\end{aligned} \tag{26}$$

The generalized momentum vector $P(t)$ is:

$$P^{\mathrm{T}}(t) = \begin{bmatrix} p_1(t), & p_2(t), & p_3(t) \end{bmatrix}. \tag{27}$$

We solve for rapid hand movements in Appendix 2, the generalized momentum vector $P(t)$ becomes:

$$P^{\mathrm{T}}(t) = \begin{bmatrix} p_1, & -p_1 t, & 2\alpha_h m \dot{F} \end{bmatrix}. \tag{28}$$

Here $p_1$ is the same as in (20). The impulse is $I(t) = \int_{-T/2}^{t} F(t')\, dt'$, and we find in Appendix 2 that the yank $\dot{F}(t)$ that gives the optimal trajectory relates to the impulse as:

$$\begin{aligned}\dot{F}(t) &= I(t) + \left(p_1 \,/\, \alpha_h m\right) t^2 \\ &\quad + \left(\dot{F}(-T/2) - \left(p_1 \,/\, \alpha_h m\right)(T/2)^2\right).\end{aligned} \quad (29)$$

The Hamiltonian $H$ may be rewritten as:

$$H = \alpha_h \cdot \left(\dot{F}^2 - \omega_h^2 F^2\right) + p_1 \cdot \left(\dot{x} - \ddot{x}t\right). \quad (30)$$

We evaluate (30) along the optimal trajectory in (21) in Appendix 3; we find:

$$H = \left(1 \,/\, 4\alpha_h \omega_h^4 m^2\right) p_1^2 + C_1 p_1 + \left(\varepsilon_h \omega_h^4 m^2\right) C_3^2. \quad (31)$$

Along the optimal trajectory, the Hamiltonian $H$ takes on a constant value $H = \Psi$. Combining (22) and (31), we find:

$$\begin{aligned}&\left(\alpha_h \,/\, 4\varepsilon_h^2 m^2\right) \psi\left(\omega_h T \,/\, 2\right) p_1^2 = \Psi, \\ &\psi\left(\omega_h T \,/\, 2\right) = 1 + \left(\omega_h T \,/\, 2\right)^2 \,/\, \sinh^2\left(\omega_h T \,/\, 2\right) \\ &\quad - 2\left(\omega_h T \,/\, 2\right) \coth\left(\omega_h T \,/\, 2\right) + \left(\omega_h T \,/\, 2\right)^2.\end{aligned} \quad (32)$$

Combining (23) and (32), we find:

$$\pi\left(\omega_h T \,/\, 2\right)^2 \,/\, \psi\left(\omega_h T \,/\, 2\right) = \left(\varepsilon_h \omega_h^4 m^2 \,/\, 4\right) D^2 \,/\, \Psi. \quad (33)$$

We observe that (33) establishes a relationship between the movement time $T$ and the generalized energy $\Psi$ so that given a value of one we can calculate the value of the other.

*4.5 Discussion*

We observed in Sec. 3 that the cost functional in (4) fails to fully accord with the two principles that we gave in Sec. 1 as we have failed provide a psychophysical account of the yank term and it is unclear to what extent a subject perceives the yank of a movement. However, we see that (29) relates the yank to the impulse. Since rapid hand movements begin and end at rest, the impulse in this case is simply the classical momentum of the hand. Thus, the relationship expressed in (29) for an optimal movement allows us to relate the yank to the movement velocity. Therefore, given a psychophysical model of the perception of the movement velocity, we may rewrite (4) in a form that satisfies the two principles in Sec. 1 but without requiring a psychophysical model of the perception of the yank. The resulting cost functional will contain two parameters that are selected by the subject for the movement as well as terms in the time $t$. The body does appear to use velocity information, and in one case of human movement, quiet standing, velocity information has been shown to provide better information for controlling the body than position or acceleration information. [42]

In practice, subjects do not select movements by selecting a specific movement time $T$, but rather have a somewhat inexact perception of the movement time $T$. As we would like a model that reflects how subjects select movements in practice, we would prefer a model in which the subject selects some other quantity which determines the movement time $T$. Moreover, we would like this quantity to have a somewhat complicated relationship to the movement time $T$ to reflect the subjects' inexact perception of the movement time $T$. We therefore suppose that the subject selects the generalized energy $\Psi$ of the movement which then takes a time $T(\Psi)$ according to (32). We have discussed this relationship between the movement time and generalized energy and how we can use the generalized energy to eliminate the need for the subject to know the movement time before executing a movement further in [14].

## 5 Rapid, Targeted Hand Movements

Rapid, targeted hand movements are rapid hand movements where the movement should end within a small but well-defined target region with high probability. To be more mathematically specific, we require that a point on the hand should end within the target region with high probability. This latter formulation of accuracy reflects the situation that we intend to analyze empirically, namely movements of a computer mouse on a tabletop. In this case, a point on the mouse corresponds to the position of the pointer on the computer screen and the aim of the movement is to have the motion of the pointer end within the area of an icon on the screen. We look at how the rapid hand movement model should be modified to produce movement times $T$ that obey Fitts' law. Moreover, we go beyond Fitts' law and provide an account of the movement accuracy in terms of the probability of hitting the target and the speed/accuracy trade-off.

*5.1 Rapid Hand Movements Approximation*

The formulas in (23) and (32) contain complicated functions of the movement time $T$. As these fumctions will complicate the derivation of Fitts' law from the model, we approximate them with functions of having a more convenient form.

Rapid hand movements have relatively low movement times $T$. We will define rapid hand movements to be those with movement times $T$ such that:

$$T < 2 / \omega_h. \tag{34}$$

We discuss this assumption further in Sec. 5.7. We use (34) to approximate $\pi(\omega_h T/2)^2/\psi(\omega_h T/2)$ to simplify the relationship between the movement time $T$ and the generalized energy $\Psi$ in (33). We approximate the $\pi(\omega\_h T/2)^2/\psi(\omega_h T/2)$ to second-order using the truncated series expansion in Appendix 4; we find:

$$\pi\left(\omega_h T / 2\right)^2 / \psi\left(\omega_h T / 2\right) \approx \frac{\left(\omega_h T\right)^6}{14400}\left(1 - \frac{\left(\omega_h T\right)^2}{70}\right). \tag{35}$$

We now use (35) to approximate (33) to second-order using the truncated series expansion in Appendix 5; we find:

$$T - \left(\omega_h^2 / 420\right)T^3 \approx \left(\left(3600\alpha_h m^2\right)D^2 / \Psi\right)^{1/6}. \tag{36}$$

For the case given in (34), we find for the second term on the right-hand side of (36) that $(\omega_h^2/420)T^3 < T/105$. Thus we can make the approximation $T - (\omega_h^2/420)T^3 \approx T$, and (36) becomes:

$$T \approx \left(\left(3600\alpha_h m^2\right) D^2 / \Psi\right)^{1/6}. \tag{37}$$

*5.2 Fitts' Law*

We take the center of the target region to lie a distance $D$ from the starting position of that point on the hand and give the target region a target width $W$ along the one-dimensional trajectory of motion of the hand. We observe that in Fitts' law in (1), the movement distance $D$ only appears as part of a ratio $D/W$ with the target width $W$. If we assume, as we did for Fitts' law in Sec. 2, that neither the movement distance $D$, nor the target width $W$ appear in the final form of (37) except as part of the ratio $D/W$, then we find that subjects select rapid, targeted hand movements having generalized energies $\Psi \sim W^2$. We write the generalized energy as:

$$\Psi \approx \left(3600\alpha_h m^2 / \tau_h^6\right) W^2. \tag{38}$$

The parameter $\tau_h$ is a constant with units of time. We introduce (38) as an ad hoc formula that produces a model that approximates Fitts' law. In Sec. 5.4, we look at how (38) arises in a utility model optimizing a speed/accuracy trade-off.

In Sec. 2, we analyzed Fitts' law using an additive model consisting of two stages: (i) a planning stage requiring a time $\tau_h$, and (ii) a motor stage requiring a time $T_M(D/W)$ to complete. The rapid hand movements model that we developed in Sec. 4 describes the motor stage of a movement. Thus, combining (37) and (38) gives the motor portion of Fitts' law:

$$T_M\left(D/W\right) \approx \tau_h \cdot \left(D/W\right)^{1/3}. \tag{39}$$

*5.3 Empirical Study (MacKenzie, 1995)*

MacKenzie [22] made empirical measurements of typical values of the parameters $a$ and $b$ in Fitts' law as given in (1) for three methods of deleting an icon on an Apple Macintosh computer. [22] The first method was *point-select*; subjects carried out point-select deletion by first selecting the icon by clicking on it and then deleting the icon by selecting the trashcan by clicking on it. The second method was *drag-select*; subjects carried out drag-select deletion by clicking on the icon, but instead of the releasing the button, holding the button down and dragging the icon to the trashcan. The third method was *stroke-through*; subjects carried out stroke-through deletion by bringing the pointer to one side of the icon, hold the button down and stroke through the icon, and finally release the mouse button on the other side of the icon. Data were obtained for ratios $D/W$ of the movement distance $D$ to the target width $W$ on the range $1.0 \leq D/W \leq 64$.

Of the three methods point-select, drag-select, and stroke-through, the case of point-select movements most resembles the model of rapid, targeted movements that we have developed. This is because we have assumed negligible friction between the mouse and the table and the drag-select and stroke-through methods require holding the mouse button down through the movement thereby increasing the friction between the mouse and the table over that present for point-select. We therefore restrict the analysis to point-select movements. For point-select movements, Fitts' law was found to take the form:

$$T = \left[230\,\text{msec}\right] + \left[166\,\text{msec/bit}\right] \log_2\left(D/W + 1\right). \tag{40}$$

Combining (2), (3), and (39), we arrive at an optimal control model of rapid, targeted hand movements that we can use to approximate (40):

$$T \approx \tau_P + \tau_h \cdot (D/W)^{1/3}. \tag{41}$$

We estimate the parameters $\tau_P$ and $\tau_h$ in (41) by minimizing the square of the difference between (40) and (41) on the range $1.0 \leq D/W \leq 64$. We find that (40) approximates (41) with $R^2 = 0.98$ and $p < 0.0001$, and parameter values:

$$\begin{aligned} \tau_P &\approx 240 \ msec, \\ \tau_h &\approx 260 \ msec. \end{aligned} \tag{42}$$

We compare the two models – (40), and the model obtained by combining (41) and (42) – in Fig. 2.

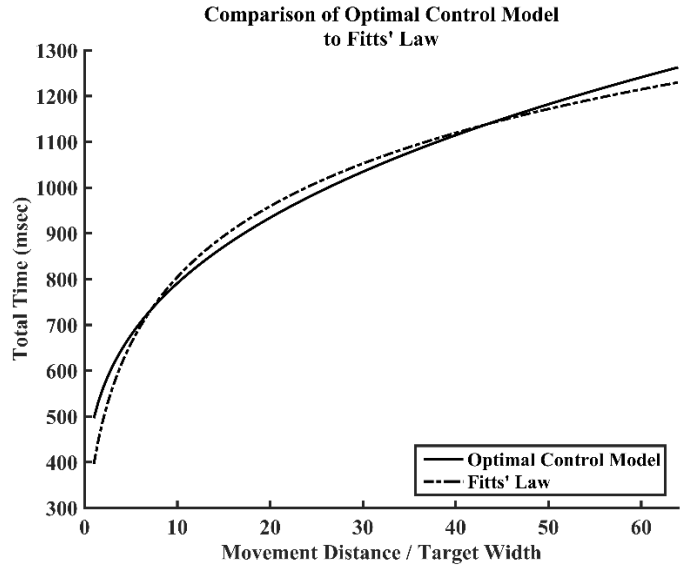

**Figure 2.** Comparison of optimal control model to Fitts' law. The optimal control model in (41) is compared to the empirical formulation of Fitts' law in (40) using the parameter values estimated in (42) over the range $0.5 \leq D/W \leq 64$ observed by MacKenzie.

*5.4 Speed/Accuracy Trade-Off*

We would like the generalized energy $\Psi$ in (38) that we used to develop a model of the motor portion of Fitts' law to result from a model of the speed/accuracy trade-off. We expect the accuracy of a movement to decrease as the movement time $T$ decreases. However, as we have already discussed, we would prefer a model in which the subject has an imperfect knowledge of the movement time $T$. Looking at (37), we note that $\Psi \sim T^{-6}$. We assume that the probability of hitting the target is a function of the generalized energy $\Psi$, and denote this by $Pr(H|\Psi)$ and suppose that the probability $Pr(H|\Psi)$ falls off approximately exponentially with the generalized energy $\Psi$, and that as $\Psi \to 0$ we have $Pr(H|\Psi) \to 1$, we find:

$$\Pr\left(H \mid \Psi\right) \approx \exp\left(-\lambda \Psi\right). \tag{43}$$

We may express the utility of executing a movement with a higher generalized energy $\Psi$ versus hitting the target with higher probability $Pr(H|\Psi)$ using a utility function of the form:

$$J = \kappa \log\left(\Psi \,/\, \Psi_0\right) + \log \Pr\left(H \mid \Psi\right). \tag{44}$$

Here $\kappa$ is a parameter whose value is selected by the subject to express the relative importance of having a shorter movement time versus having a more accurate movement, and $\Psi_0$ is a constant with units of generalized energy included to make the argument of $log(\Psi/\Psi_0)$ dimensionless. The subject selects the rapid, targeted hand movement by maximizing the utility $J$ in (44). We have adopted the language of a utility model at this point to describe a higher level selection of rapid, targeted hand movements here to parallel the use of a utility model to describe the selection of a walking gait in [5]. In effect, we have used the language of cost in [6] and in this paper to express a lower level optimization of the movement trajectory.

Inserting (43) into (44) gives:

$$J \approx \kappa \log\left(\Psi \,/\, \Psi_0\right) - \lambda \Psi. \tag{45}$$

We maximize the utility $J$ by taking the derivative of the utility function in (45) with respect to the generalized energy $\Psi$ and setting it to zero:

$$dJ \,/\, d\Psi \approx \kappa \,/\, \Psi - \lambda = 0. \tag{46}$$

Solving (46) for the selected generalized energy $\Psi$, we find:

$$\Psi \approx \kappa \,/\, \lambda. \tag{47}$$

Combining (38) and (47), we find the ratio of parameter values $\kappa/\lambda$ that cause the generalized energy $\Psi$ in (38) to arise from the model of the speed/accuracy trade-off to be:

$$\kappa \,/\, \lambda \approx \left(3600 \alpha_h m^2 \,/\, \tau_h^6\right) W^2. \tag{48}$$

We assume that all the information related to the speed of the movement (i.e. $\tau_h$) to appear as part of $\kappa$, while all the information about the target (i.e. the movement distance $D$ and the target width $W$) to appear as part of $\lambda$ which is the representative of the probability $Pr(H|\Psi)$ of hitting the target. Introducing a constant parameter $\tau_A$, we may solve for $\kappa$ and $\lambda$ in (48):

$$\begin{aligned}\kappa &\approx \left(\tau_A \,/\, \tau_h\right)^6, \\ \lambda &\approx \left(\tau_A^6 \,/\, 3600 \alpha_h m^2\right) W^{-2}.\end{aligned} \tag{49}$$

Inserting (49) into (43) gives the probability $Pr(H|\Psi)$ of hitting the target consistent with the generalized energy $\Psi$ in (38):

$$\Pr\left(H \mid \Psi\right) \approx \exp\left(-\left(\tau_A^6 \,/\, 3600 \alpha_h m^2\right) \Psi \,/\, W^2\right). \tag{50}$$

Inserting (38) into (50), we may rewrite the probability $Pr(H|\Psi)$ of hitting the target given in (50) in terms of the parameter value $\tau_h$:

$$\Pr\left(H \mid \tau_h\right) \approx \exp\left(-\left(\tau_A / \tau_h\right)^6\right). \tag{51}$$

The probability $Pr(H|\Psi)$ of hitting the target given in (50) and $Pr(H|\tau_h)$ given in (51) is valid for movements in general. This means the subject may resolve the speed/accuracy trade-off in an arbitrary way for each movement, or, put another way, the subject may select an arbitrary value for $\tau_h$ for each movement. However, for a set of movements made under the same conditions we expect the resolution of speed/accuracy trade-off and the resulting $\tau_h$ to be fixed for all the movements. The solution to the speed/accuracy trade-off is determined by the value of the parameter κ in (44). The subject gives the relative weighting of speed versus accuracy by choosing κ. We find that this parameter is related to the probability $Pr(H|\tau_h)$ of hitting the target according to:

$$\kappa \approx -\log \Pr\left(H \mid \tau_h\right) \tag{52}$$

Thus, the selection of κ amounts to a selection of the probability $Pr(H|\tau_h)$ of hitting the target and this probability is independent of the movement distance $D$ and the target width $W$.

*5.5 Empirical Study (Seelye et al. 2015)*

Seelye et al. [30] examined the routine computer usage of a cohort of 42 intact older adults (age = 88 ± 5.2 years, 37 female / 5 male) and 20 older adult with mild cognitive impairment (MCI) (age = 88 ± 6.6 years, 16 female / 4 male) using data taken from a single one week period for each subject. The computer movement data were obtained and analyzed and characterized using several metrics, but the metric of interest to us is the total distance moved by the mouse for each movement (denoted by $D$ in Seelye et al.). This distance was measured in counts where counts are a representation of the distanced moved that is internal to the computer, and while closely related to both measurements in pixels on the display and distances moved by the mouse along the tabletop. In their analysis, Seelye et al. characterize the distribution of distances moved for observed mouse movements for each subject using the median of the observed distances which we denote $M\{D\}$, and then characterize intact and MCI subgroups of the cohort using the mean of the medians which we denote $\langle M\{D\}\rangle$. The observed values for this were $\langle M\{D\}\rangle_{Intact} = 57 \pm 25$ counts and $\langle M\{D\}\rangle_{MCI} = 42 \pm 19$ counts. The numbers of movements available from each subgroup for analysis were $N_{Intact} = 7900 \pm 9700$ and $N_{MCI} = 1500 \pm 1700$.

We analyze the data reported by Seelye et al. by assuming that the observed difference in $\langle M\{D\}\rangle$ between the two subgroups is due entirely to a difference in the probability of hitting target icons and the consequent additional movement needed to hit the target icons after a miss. We discuss this assumption further in Sec. 5.7. As we have shown in Sec. 5.4, the subject selects the speed/accuracy trade-off by selecting the probability $Pr(H)$ of hitting the target, and this probability is independent of the movement distance $D$ or the target of width $W$. We construct a simple movement error model that assumes that the subject moves along a straight line to the target and misses the target either by stopping short of the target or going beyond the target. We further assume the subject always misses the target by a distance $\delta D$, that the subject stops short of the target with probability $p$ and goes beyond the target with probability $1 - p$, and that when the subject misses a target the subject immediately makes a short move of distance $\delta D$ to the missed target and always hits the target with the next movement. Using this model, we find that, for two populations $A$ and $B$ observed to have average movement distances $\langle D_{move}\rangle_A$ and $\langle D_{move}\rangle_B$, the probabilities $Pr(M|A)$ and $Pr(M|B)$ approximately related to each other as (see Appendix 6):

$$\Pr(M \mid A) \approx \frac{\langle D_{move}\rangle_B - \langle D_{move}\rangle_A}{\langle D_{move}\rangle_A} + \frac{\langle D_{move}\rangle_B}{\langle D_{move}\rangle_A}\Pr(M \mid B). \tag{53}$$

Under the assumption that this model holds approximately, we may therefore estimate the relationship between the probabilities $Pr(M|A)$ and $Pr(M|B)$ of the two groups by observing the statistics of the distances moved by the mouse. Taking $\langle D \rangle$ in the model in (53) to be approximately equal to $\langle M\{D\}\rangle$ in the data, we then estimate:

$$\Pr(M \mid MCI) \approx 0.36 + 1.4\Pr(M \mid Intact). \tag{54}$$

As we expect $Pr(M|Intact)$ to be low, we estimate that an average subject in the MCI subgroup misses the target about one third of the time.

*5.6 Maximizing the Utility Function in Practice*

While the formal mathematical procedure of maximizing the utility function $J$ in (44) in Sec. 5.4 proceeds neatly, it remains to consider how a subject might carry out this maximization in practice. We begin the process of constructing a practical model of the maximization by defining a random variable (RV) $\psi_\Psi^\kappa$ that satisfies the probabilities:

$$\begin{aligned}\Pr\left(\psi_\Psi^\kappa = (\Psi/\Psi_0)^\kappa\right) &= \Pr(H \mid \Psi),\\ \Pr\left(\psi_\Psi^\kappa = 0\right) &= \Pr(M \mid \Psi).\end{aligned} \tag{55}$$

Thus, the RV $\psi_\Psi^\kappa$ returns a value $(\Psi/\Psi_0)^\kappa$ with probability $Pr(H|\Psi)$ and return zero otherwise. Denoting the expected value of $\psi_\Psi^\kappa$ by $E\{\psi_\Psi^\kappa\}$, we can replace the utility function $J$ in (44) with an equivalent utility function $J'$ given by:

$$J' = e^J = E\left\{\psi_\Psi^\kappa\right\}. \tag{56}$$

It should be clear that the value $\Psi$ that maximizes $J'$ also maximizes $J$ and that the generalized energy $\Psi$ that maximizes the utility function in the speed/accuracy trade-off is the one that maximizes the expected value of the RV $\psi_\Psi^\kappa$.

It is now possible to provide a somewhat inefficient procedure by which the subject may estimate the expectation $E\{\psi_\Psi^\kappa\}$ for a generalized energy $\Psi$ without having to have an explicit model for the probability $Pr(H|\Psi)$ of hitting the target. For a given target width $W$, the subject may proceed by sampling multiple movements and scoring each movement that hits the target with a value $(\Psi/\Psi_0)^\kappa$ and scoring each that misses the target with a zero, and then take the mean of the scores. By repeating this process over a variety of conditions, the subject will arrive at the desired maximum. As we have said, this procedure is somewhat inefficient, and we would expect subjects to adopt a more efficient search procedure built around sampling movements and making modifications to how the movements are selected that is expected to maximize to find movements that maximize the expectation $E\{\psi_\Psi^\kappa\}$. However, the exploration of such procedures is beyond the scope of this paper.

*5.7 Discussion*

We have developed an optimal control model in Secs. 3 and 4 that contains both jerk and force terms, the approximations we have made in this section to yield the expression in (41) have effectively taken $\varepsilon_h \approx 0$ and used a model only involving the jerk term. We are effectively assuming that $\alpha_h$ is sufficiently large compared to $\varepsilon_h$ given the typical magnitudes of the yanks and forces generated by the muscles that we expect to see in practice that the cost associated with the jerk is much larger than the cost associated with the forces. This is the same approximation that we made for the optimal control model for a step during walking gait in [6]. We may put an approximate bound on the relationship between $\alpha_h$ and $\varepsilon_h$ by assuming that the largest value of $D/W$ observed by MacKenzie [22] is approximately the limit of applicability of the rapid hand movements approximation in Sec. 5.1. The largest value is about $D/W = 64$ giving the largest movement times of around 1300 msec using the model in (41) with parameter values in (42). This gives a bound of about $\omega_h < 1.5$ Hz or $\alpha_h > [0.42 \text{ sec}^2] \varepsilon_h$.

We assumed a simple exponential model of movement accuracy in (43) in terms of the generalized energy $\Psi$. We argued for this by noting the relationship between the generalized energy $\Psi$ and the movement time $T$ in (37) and noting that we wanted a model in which the movement accuracy increased with movements taking a longer time $T$. We find the resulting model has the qualitative behavior we would expect intuitively, namely a greater movement accuracy for movements taking a longer time $T$ and a lower movement accuracy for movements with a larger ratio $D/W$. We should think of the simple exponential model of movement accuracy as an approximate model that describes a range of rapid, targeted, movements of the hand that are typical and comfortable, and breaks down for movements outside of the typical range. Examples of such movements would include very fast or very slow movements, movements with very large ratios $D/W$, or movements with approach physical limits governing hand movements.

We have made a rough comparison of computer mouse movement during routine computer usage between a group of intact subjects and a group of subjects with MCI. By characterizing each group by the difference between the observed average of the median movement distance for its members and assuming that the observed differences were entirely due to differences in movement accuracy, we found that the MCI group typically missed the target one third of the time. While we do expect the movement accuracy to affect the observed mouse movement statistics it is certainly possible that other behaviors exhibited by the MCI group contribute to the differing observed statistics. One possibility that we did not include in the model would be that the MCI group makes multiple shorter movements where the intact group makes a single movement. However, this and other possibilities would amount to a variety of strategies for dealing with a lower movement accuracy, and thus we would expect models accounting for such strategies to provide a more sophisticated version of the analysis given in Sec. 5.5, but nevertheless allowing for the estimation of the probability of missing the target given the observed movement data.

**Appendix 1**

The trajectory $x(t)$ is given by:

$$x(t) = C_1 t - \left( p_1 \, / \, 12\varepsilon_h m^2 \right) t^3 + C_3 \sinh(\omega_h t). \tag{57}$$

This trajectory must satisfy the initial and final conditions:

$$\begin{aligned}
x(-T/2) &= -D/2, & x(T/2) &= D/2, \\
\dot{x}(-T/2) &= 0, & \dot{x}(T/2) &= 0, \\
\ddot{x}(-T/2) &= 0, & \ddot{x}(T/2) &= 0.
\end{aligned} \tag{58}$$

Combining (57) and (58) yields the system of three equations in three unknowns given by:

$$\begin{bmatrix} T/2 & \sinh(\omega_h T/2) \\ 1 & \omega_h \cosh(\omega_h T/2) \\ 0 & \omega_h^2 \sinh(\omega_h T/2) \end{bmatrix} \begin{bmatrix} C_1 \\ C_3 \end{bmatrix} = \begin{bmatrix} D/2 + p_1 T^3/96\varepsilon_h m^2 \\ p_1 T^2/16\varepsilon_h m^2 \\ p_1 T/4\varepsilon_h m^2 \end{bmatrix}. \tag{59}$$

We can immediately read the solutions to $C_1$ and $C_3$ from (59) using the equations given in the second and third rows; they are:

$$\begin{aligned}
C_1 &= \left(1/4\alpha_h \omega_h^4 m^2\right)\left((\omega_h T/2)^2 - 2(\omega_h T/2)\coth(\omega_h T/2)\right) p_1, \\
C_3 &= \left(1/2\alpha_h \omega_h^5 m^2\right)\left((\omega_h T/2)/\sinh(\omega_h T/2)\right) p_1.
\end{aligned} \tag{60}$$

Combining the first row in (59) with the values for $C_1$ and $C_3$ given in (60), we find:

$$\begin{aligned}
p_1 &= \left(\alpha_h \omega_h^5 m^2\right) D / \pi(\omega_h T/2), \\
\pi(\omega_h T/2) &= -(\omega_h T/2)^2 \left(\coth(\omega_h T/2) - (\omega_h T/2)^{-1} - (\omega_h T/2)/3\right).
\end{aligned} \tag{61}$$

**Appendix 2**

The generalized momentum vector $P(t)$ is:

$$P^{\mathrm{T}}(t) = \begin{bmatrix} p_1(t), & p_2(t), & p_3(t) \end{bmatrix}. \tag{62}$$

Using equation C in (9), we find the optimal trajectory satisfies $\partial H/\partial u = 0$; this gives:

$$\begin{aligned}
p_3(t) &= 2\alpha_h m^2 u \\
&= 2\alpha_h m \dot{F}.
\end{aligned} \tag{63}$$

Thus, the generalized momentum vector $P(t)$ becomes:

$$P^{\mathrm{T}}(t) = \begin{bmatrix} p_1(t), & p_2(t), & 2\alpha_h m\dot{F} \end{bmatrix}. \tag{64}$$

Using equation B in (9), we find that the generalized momentum vector $P(t)$ for optimal trajectory satisfies $\dot{P} = -\partial H/\partial Q$; this gives:

$$\dot{P}(t) = -\begin{bmatrix} 0 & 0 & 0 \\ 1 & 0 & 0 \\ 0 & 1 & 0 \end{bmatrix} P(t) + 2\varepsilon_h m \begin{bmatrix} 0 \\ 0 \\ F(t) \end{bmatrix}. \tag{65}$$

Therefore, the generalized momentum vector $P(t)$ is:

$$P(t) = 2\alpha_h m \begin{bmatrix} \left(\ddot{F} - \omega_h^2 F\right) \\ -\left(\dddot{F} - \omega_h^2 \dot{F}\right) \\ \dot{F} \end{bmatrix}. \tag{66}$$

We can use the solution for the force applied to the hand during an optimal movement of the hand (24) to simplify the first and second components of the generalized momentum vector $P(t)$ in (66); we find:

$$\begin{aligned} \ddot{F} - \omega_h^2 F &= \left(p_1 / 2\alpha_h m\right) t, \\ \dddot{F} - \omega_h^2 \dot{F} &= \left(p_1 / 2\alpha_h m\right). \end{aligned} \tag{67}$$

We note that the value $p_1$ is the constant parameter defined in (20) and appearing in the optimal trajectory of the hand in (21). Combining (66) and (67), we find:

$$P^{\mathrm{T}}(t) = \begin{bmatrix} p_1, & -p_1 t, & 2\alpha_h m \dot{F} \end{bmatrix}. \tag{68}$$

Taking the first equation in (67), integrating over time, and recalling that the impulse is $I(t) = \int_{-T/2}^{t} F(t')\, dt'$, we find the yank $\dot{F}(t)$ that gives the optimal trajectory is:

$$\begin{aligned} \dot{F}(t) &= I(t) + \left(p_1 / \alpha_h m\right) t^2 \\ &\quad + \left(\dot{F}(-T/2) - \left(p_1 / \alpha_h m\right)(T/2)^2\right). \end{aligned} \tag{69}$$

**Appendix 3**

The Hamiltonian $H$ of the system is given by:

$$H = \alpha_h \cdot \left(\dot{F}^2 - \omega_h^2 F^2\right) + p_1 \cdot \left(\dot{x} - \ddot{x} t\right). \tag{70}$$

The force $F$ and yank $\dot{F}$ are given by:

$$\begin{aligned} F &= -\left(p_1 / 2\varepsilon_h m\right) t + C_3 \omega_h^2 m \sinh(\omega_h t), \\ \dot{F} &= -\left(p_1 / 2\varepsilon_h m\right) + C_3 \omega_h^3 m \cosh(\omega_h t). \end{aligned} \tag{71}$$

The velocity $\dot{x}$ and acceleration $\ddot{x}$ are given by:

$$\begin{aligned} \dot{x} &= C_1 - 3\left(p_1 / 12\varepsilon_h m^2\right) t^2 + C_3 \omega_h \cosh(\omega_h t), \\ \ddot{x} &= -6\left(p_1 / 12\varepsilon_h m^2\right) t + C_3 \omega_h^2 \sinh(\omega_h t). \end{aligned} \tag{72}$$

We begin by calculating expressions for the force factors appearing the Hamiltonian in (70). Taking the squares of the force $F$ and the jerk $\dot{F}$ in (71), we find:

$$\begin{aligned} F^2 &= \left(p_1 / 2\varepsilon_h m\right)^2 t^2 - C_3 \omega_h^2 \left(p_1 / \varepsilon_h\right) t \sinh(\omega_h t) \\ &\quad + C_3^2 \omega_h^4 m^2 \sinh^2(\omega_h t), \\ \dot{F}^2 &= \left(p_1 / 2\varepsilon_h m\right)^2 - C_3 \omega_h^3 \left(p_1 / \varepsilon_h\right) \cosh(\omega_h t) \\ &\quad + C_3^2 \omega_h^6 m^2 \cosh^2(\omega_h t). \end{aligned} \tag{73}$$

We can now calculate the factor $\dot{F}^2 - \omega_h^2 F^2$ in the Hamiltonian in (70) using the squares of the force $F$ and the jerk $\dot{F}$ in (73); we find:

$$\begin{aligned}\dot{F}^2 - \omega_h^2 F^2 &= C_3^2 \omega_h^6 m^2 + \left(p_1 / 2\varepsilon_h m\right)^2 \left(1 - \left(\omega_h t\right)^2\right) \\ &\quad - C_3 \omega_h^3 \left(p_1 / \varepsilon_h\right)\left(\cosh\left(\omega_h t\right) - \left(\omega_h t\right)\sinh\left(\omega_h t\right)\right).\end{aligned} \quad (74)$$

We next calculate a useful expression for the velocity $\dot{x}$ and acceleration $\ddot{x}$ factors appearing in the Hamiltonian in (70). Using (72), we calculate $\dot{x} - \ddot{x}t$ to be:

$$\begin{aligned}\dot{x} - \ddot{x}t &= C_1 + \left(p_1 / 4\alpha_h \omega_h^4 m^2\right)\left(\omega_h t\right)^2 \\ &\quad + C_3 \omega_h \cdot \left(\cosh\left(\omega_h t\right) - \left(\omega_h t\right)\sinh\left(\omega_h t\right)\right).\end{aligned} \quad (75)$$

Combining (70), (74), and (75), we calculate the Hamiltonian $H$ to be:

$$H = \left(1 / 4\alpha_h \omega_h^4 m^2\right) p_1^2 + C_1 p_1 + \left(\varepsilon_h \omega_h^4 m^2\right) C_3^2. \quad (76)$$

**Appendix 4**

We calculate series expansions truncated to the lowest two orders for the functions:

$$\begin{aligned}\pi(\xi) &= -\xi^2 \left(\coth \xi - \xi^{-1} - \xi/3\right), & (A) \\ \psi(\xi) &= 1 + \xi^2 / \sinh^2 \xi - 2\xi \coth \xi + \xi^2. & (B)\end{aligned} \quad (77)$$

We do this using the following three truncated series expansions:

$$\begin{aligned}(1-\xi)^{-1} &\approx 1 + \xi + \xi^2 + \xi^3, & (A) \\ \sinh \xi &\approx \xi + \frac{1}{6}\xi^3 + \frac{1}{120}\xi^5 + \frac{1}{5040}\xi^7 & (B) \\ \coth \xi &\approx \xi^{-1} + \frac{1}{3}\xi - \frac{1}{45}\xi^3 + \frac{2}{945}\xi^5 - \frac{1}{4725}\xi^7. & (C)\end{aligned} \quad (78)$$

We note that equation A in (78) requires the assumption that $|\xi| < 1$.

We first approximate Eq. A in (77). Inserting Eq. C in (78) into Eq. A in (77) and simplifying, we find:

$$\pi(\xi) \approx \frac{1}{45}\xi^5 \left(1 - \frac{2}{21}\xi^2\right). \quad (79)$$

We next develop a truncated series expansion for the second term of Eq. B in (77). Using Eq. B in (78), we find:

$$\frac{\xi^2}{\sinh^2 \xi} \approx \left(1 + \frac{1}{6}\xi^2 + \frac{1}{120}\xi^4 + \frac{1}{5040}\xi^6\right)^{-2}. \quad (80)$$

We proceed by calculating the square of the argument on the right-hand side in (80). As we will only be retaining the series expansion to sixth-order in $\xi$, we can eliminate some terms at this stage that will not affect the final approximation; we find:

$$\frac{\xi^2}{\sinh^2 \xi} \approx \left[1 + 2\left(\frac{1}{6}\xi^2 + \frac{1}{120}\xi^4 + \frac{1}{5040}\xi^6\right) + \left(\frac{1}{6}\xi^2 + \frac{1}{120}\xi^4\right)^2\right]^{-1}. \tag{81}$$

Simplifying (81), and eliminating further terms that will not affect the final approximation, we find:

$$\frac{\xi^2}{\sinh^2 \xi} \approx \left[1 + \frac{1}{3}\xi^2 + \frac{2}{45}\xi^4 + \frac{1}{315}\xi^6\right]^{-1}. \tag{82}$$

We now combine (81) with Eq. A in (78) and eliminate terms that will not affect the final approximation to obtain:

$$\begin{aligned}\frac{\xi^2}{\sinh^2 \xi} &\approx 1 - \left(\frac{1}{3}\xi^2 + \frac{2}{45}\xi^4 + \frac{1}{315}\xi^6\right) \\ &+ \left(\frac{1}{3}\xi^2 + \frac{2}{45}\xi^4\right)^2 - \left(\frac{1}{3}\xi^2\right)^3.\end{aligned} \tag{83}$$

Simplifying (83), and eliminating further terms that will not affect the final approximation, we arrive at the desired truncated series expansion:

$$\frac{\xi^2}{\sinh^2 \xi} \approx 1 - \frac{1}{3}\xi^2 + \frac{1}{15}\xi^4 - \frac{2}{189}\xi^6. \tag{84}$$

We next develop a truncated series expansion for the third term of Eq. B in (77). Using Eq. C in (78), we find:

$$\xi \coth \xi \approx 1 + \frac{1}{3}\xi^2 - \frac{1}{45}\xi^4 + \frac{2}{945}\xi^6. \tag{85}$$

Combining Eq. B in (77) with (84) and (85), and simplifying, we obtain the desired truncated series expansion:

$$\psi(\xi) \approx \frac{1}{9}\xi^4\left(1 - \frac{14}{105}\xi^2\right). \tag{86}$$

We finally calculate a truncated series expansion for the quantity $\pi(\xi)^2/\psi(\xi)$. Combining (79), (86) with Eq. A in (78) truncated to the first-order in $\xi$, we find:

$$\pi(\xi)^2 / \psi(\xi) \approx \frac{1}{225}\xi^6\left(1 - \frac{4}{21}\xi^2\right)\left(1 + \frac{14}{105}\xi^2\right). \tag{87}$$

Simplifying (87), and eliminating further terms that will not affect the final approximation, we arrive at the desired truncated series expansion:

$$\pi(\xi)^2 / \psi(\xi) \approx \frac{1}{225}\xi^6\left(1 - \frac{2}{35}\xi^2\right). \tag{88}$$

**Appendix 5**

The general relationship between movement time $T$ and movement distance $D$ for rapid movements of the hand with generalized energy $\Psi$ is given in (33); it is:

$$\pi(\omega_h T / 2)^2 / \psi(\omega_h T / 2) = \left(\varepsilon_h \omega_h^4 m^2 / 4\right) D^2 / \Psi. \tag{89}$$

The approximation of LHS of (89) to second-order using a truncated series expansion is given in (35); it is:

$$\pi(\omega_h T / 2)^2 / \psi(\omega_h T / 2) \approx \frac{(\omega_h T)^6}{14400}\left(1 - \frac{(\omega_h T)^2}{70}\right). \tag{90}$$

Combining (89) and (90), we find:

$$T \cdot \left(1 - \left(\omega_h^2 / 70\right) T^2\right)^{1/6} \approx \left(\left(3600 \alpha_h m^2\right) D^2 / \Psi\right)^{1/6}. \tag{91}$$

For $|\xi| < 1$, we can make the following approximation:

$$(1 - \xi)^{1/6} \approx 1 - \xi / 6. \tag{92}$$

Combining (91) and (92), we find:

$$T - \left(\omega_h^2 / 420\right) T^3 \approx \left(\left(3600 \alpha_h m^2\right) D^2 / \Psi\right)^{1/6}. \tag{93}$$

**Appendix 6**

We imagine a subject makes a large number $N$ of moves to a sequence of targets with widths $W_n$ at distances $D_n$ where the subject has always selected the same probability $Pr(H)$ of hitting the target, which we also right as the probability $Pr(M) = 1 - Pr(H)$ of missing the target. The expected total distance $E\{D_{total}\}$ moved by the hand is:

$$\begin{aligned} E\{D_{total}\} &= \left(\sum_{n=1}^{N} D_n\right) \Pr(H) \\ &+ \left(\left(\sum_{n=1}^{N} D_n\right) - N\delta D p + N\delta D \cdot (1 - p) + N\delta D\right) \Pr(M) \\ &= \left(\sum_{n=1}^{N} D_n\right) + 2N\delta D \cdot (1 - p) \Pr(M). \end{aligned} \tag{94}$$

The expected total number of moves $E\{N_{total}\}$ made by the subject is:

$$\begin{aligned} E\{N_{total}\} &= NP(H) + 2N\Pr(M) \\ &= N \cdot (1 + \Pr(M)) \end{aligned} \tag{95}$$

For large $N$, we expect the distributions of $D_{total}$ and $N_{total}$ to be normal around the expected values with very small standard deviations, so we can approximate the average movement distance $\langle D_{move} \rangle$ by $E\{D_{total}\}/E\{N_{total}\}$. We define the average distance to a target to be $\langle D \rangle = (\sum_{n=1}^{N} D_N)/N$, then we find:

$$\langle D_{move} \rangle \approx \frac{\langle D \rangle + 2\delta D \cdot (1-p)\Pr(M)}{(1 + \Pr(M))} \qquad (96)$$

Under the assumptions that $\delta D$ is small compared to $\langle D \rangle$ and the probability $Pr(M)$ of missing the target is small, we find:

$$\langle D_{move} \rangle \approx \langle D \rangle / (1 + \Pr(M)). \qquad (97)$$

For two groups $A$ and $B$ carrying out tasks in which $\langle D \rangle$ is approximately the same, we find:

$$\frac{\langle D_{move} \rangle_A}{\langle D_{move} \rangle_B} \approx \frac{1 + \Pr(M \mid B)}{1 + \Pr(M \mid A)}. \qquad (98)$$

Solving (98) for $Pr(M|A)$:

$$\Pr(M \mid A) \approx \frac{\langle D_{move} \rangle_B - \langle D_{move} \rangle_A}{\langle D_{move} \rangle_A} + \frac{\langle D_{move} \rangle_B}{\langle D_{move} \rangle_A} \Pr(M \mid B). \qquad (99)$$